\documentclass{jltp}

\usepackage{graphicx} % uncomment this line to include the graphicx package

\title{Decay of quantised vorticity by sound emission}

\author{C. F. Barenghi$^1$, N.G. Parker$^2$, N.P. Proukakis$^2$ and
C.S. Adams$^2$}

\address{$^1$School of Mathematics, University of Newcastle,\\
Newcastle upon Tyne, NE1 7RU, UK\\
$^2$ Department of Physics, University of Durham,
\\ Durham, DH1 3LE, UK}

\runninghead{C.F. Barenghi \textit{et al.}}{Decay of quantised vorticity}

\begin{document}

\maketitle

\begin{abstract}
It is thought that in a quantum fluid sound generation is the
ultimate sink of turbulent kinetic energy in the absence of any
other dissipation mechanism near absolute zero. We show that a
suitably trapped Bose-Einstein condensate provides a model system
to study the sound emitted by accelerating vortices in a
controlled way.

PACS numbers: 03.75.Lm, 47.32.Cc, 67.40.Vs.
\end{abstract}

\section{BACKGROUND}

If quantised vorticity is created in helium~II at temperatures so
small that the normal fluid fraction (hence viscosity and
friction) is negligible, it is found that it rapidly
decays\cite{Davis}. It is thought that the mechanism which is
responsible for kinetic energy dissipation is the creation of
sound waves, or phonons\cite{Samuels,Vinen-Niemela}. It was Nore
{\it et al.}\cite{Nore} who first found that during the evolution
of a vortex tangle the total kinetic energy decreases and the
total sound energy increases. By doing a more detailed analysis we
revealed that a short sound pulse is emitted at each vortex
reconnection\cite{Leadbeater1}. The importance of vortex
reconnections was highlighted when it was discovered
that\cite{Kivotides} the vortex cusp which is created at each
reconnection event relaxes and triggers large amplitude Kelvin
waves (helical displacements of the vortex core); these waves
interact non-linearly and generate more waves of higher and higher
wavenumber. This Kelvin wave
cascade\cite{Vinen-Tsubota,Svistunov}, which is analogous to the
Kolmogorov cascade of classical turbulence, thus shifts the
kinetic energy to wavenumbers which are large enough that a vortex
line can efficiently radiate sound\cite{Vinen}. Recently we
demonstrated\cite{Leadbeater2} the simultaneous occurrence of both
processes (reconnection pulses and sound radiation).

Unfortunately the study of quantised vorticity in helium~II
suffers from a lack of direct visualisation. Better detection
techniques exist for trapped weakly-interacting atomic
Bose-Einstein condensates (BECs). The key ingredients of the
problem (quantised vorticity, sound waves, reconnections) are
present in both systems. The aim of this paper is to show that a
trapped Bose-Einstein condensate provides a model system to study
sound radiation by vortices in a controlled way. Sound radiation
in a BEC is thus not only interesting {\it per se}, but it can
give insight into the more difficult problem of superfluid
turbulence.

\begin{figure}
\centering \includegraphics[height=1.7in,angle=0]{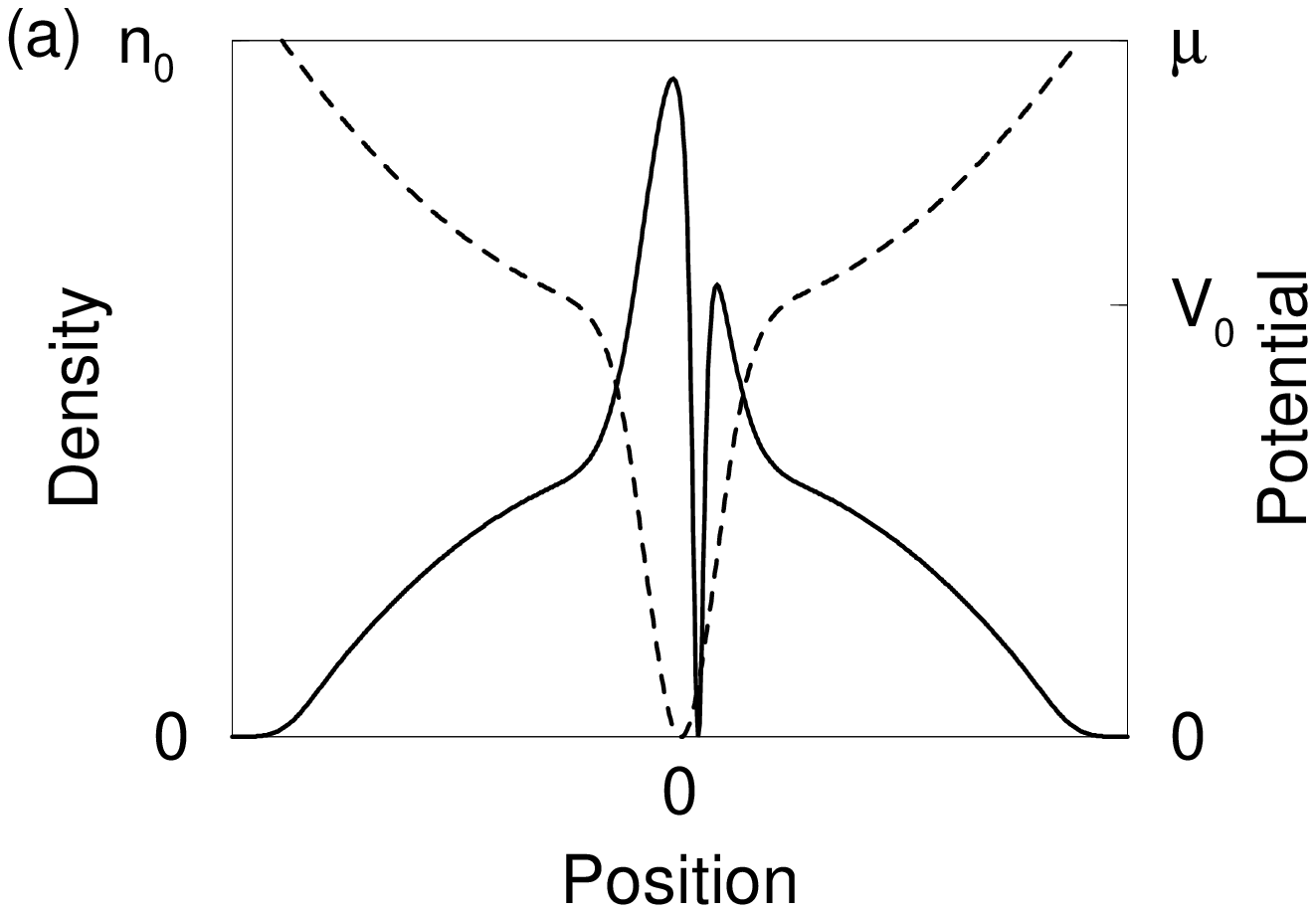}
\includegraphics[height=1.6in,angle=0]{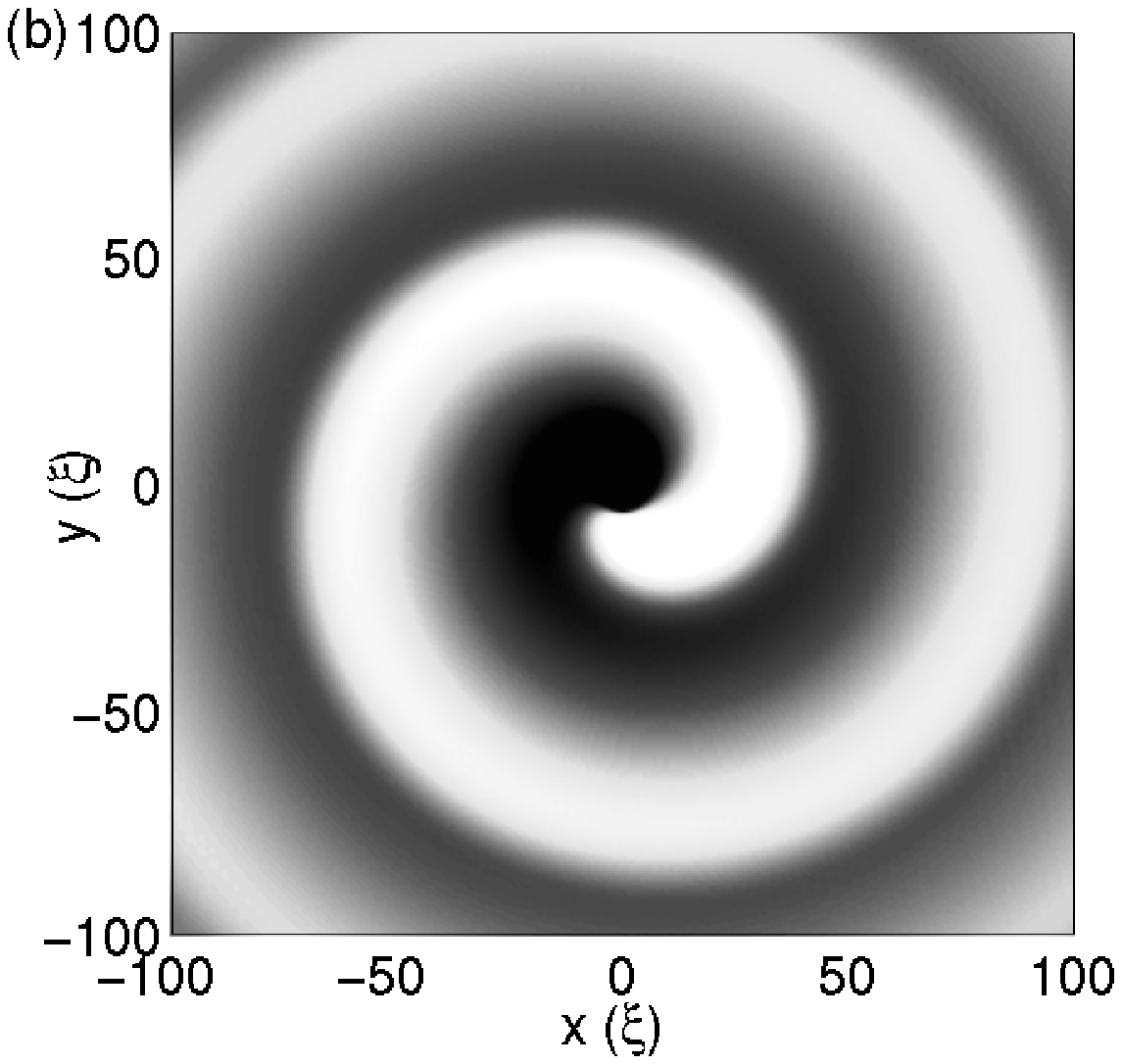} \caption{
(a) Radial density distribution $n$ (solid line), including an
off-centre vortex, for a BEC confined by a tight Gaussian dimple
in a weaker harmonic trap.
 (b) Renormalised fluid density (actual
density minus inhomogeneous background density) for $V_0<\mu$
reveals the emission of dipolar sound waves from the accelerating
vortex, with amplitude $\sim 1\% ~ n_0$.} \label{trap}
\end{figure}

\begin{figure}
\centering
\includegraphics[height=1.72in,angle=0]{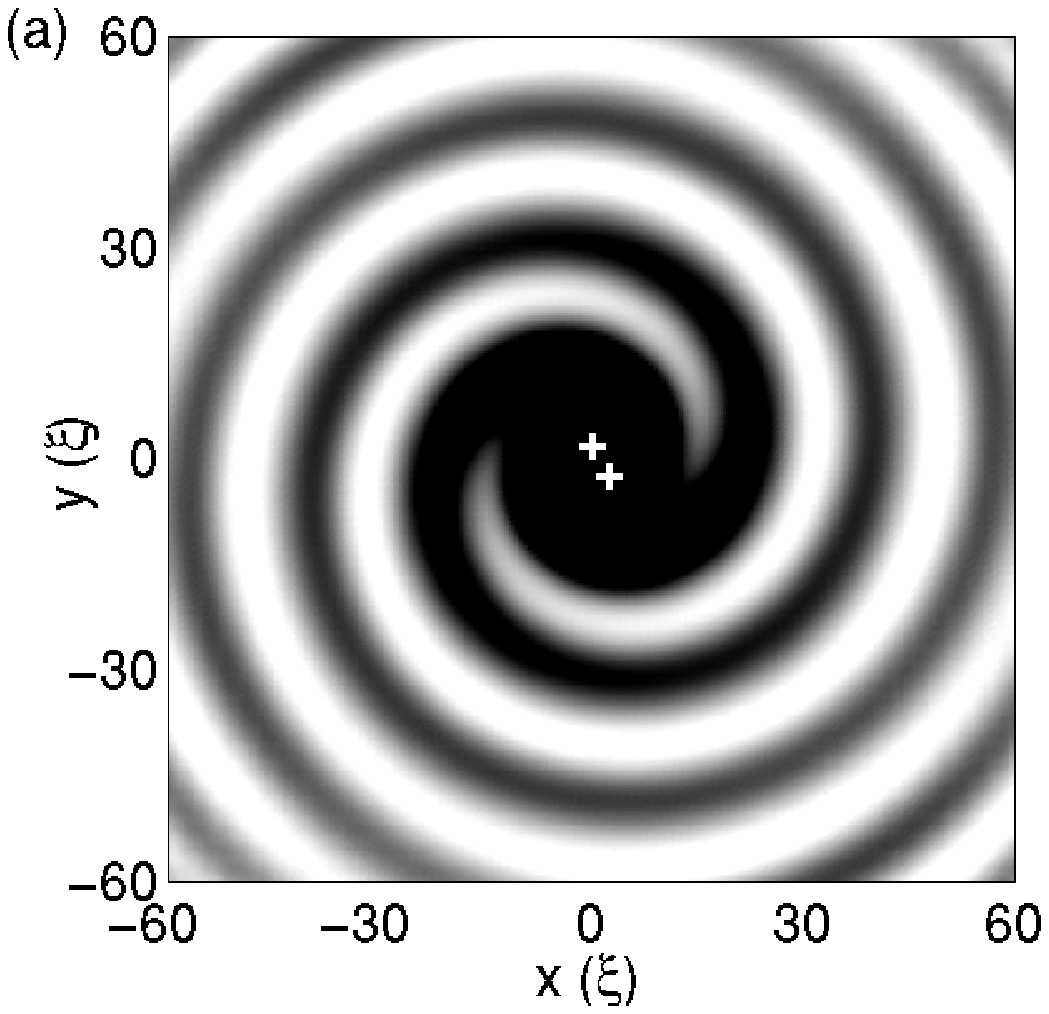}
\includegraphics[height=1.72in,angle=0]{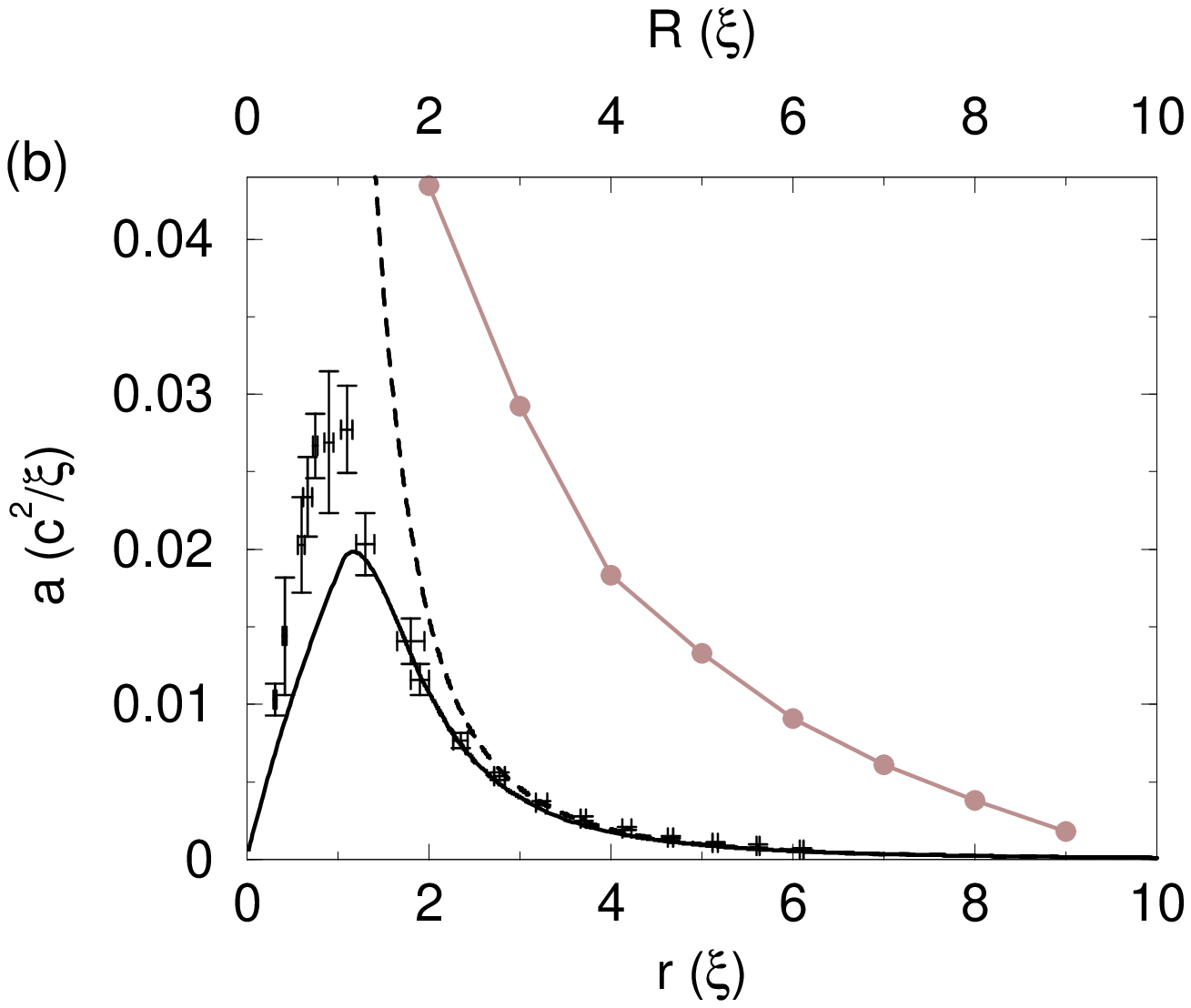}
\caption{(a) Density plot of a corotating vortex-vortex pair in a
homogeneous system.  Quadrupolar sound emission, of amplitude
$\sim 0.2\% ~n_0$, is generated from the corotating vortex cores
(illustrated by plus signs).(b) Acceleration of the pair as a
function of the radius $r$ (bottom axis), calculated from the
time-dependent GP equation (black points with error bars), a
time-independent numerical approach in the rotating frame (solid
black line), and the analytic prediction\cite{Pismen}
corresponding to $a=\kappa^2/r^3$ (dashed line). The acceleration
of a single vortex precessing in a harmonic trap $\omega=\sqrt{2}
\times 10^{-1} (\mu/\hbar)$ is also shown (grey line), as a
function of distance from the condensate edge (top axis)
$R(\xi)=R_{C}-r(\xi)$, where the condensate radius $R_{C}=10\xi$.
} \label{pair}
\end{figure}

\section{MODEL}

In a recent paper\cite{Parker} we have suggested that sound
emission by a quantised vortex can be studied in a controlled way
by letting a vortex precess within a dimple embedded in a weaker
harmonic potential which confines a quasi-two-dimensional
(quasi-2D) atomic condensate, as illustrated in Fig.~1(a).  If the
dimple depth $V_0$ is less than the chemical potential $\mu$ the
sound (which has an energy of the order of $\mu$) escapes,
otherwise it remains in the region near the vortex and can be
reabsorbed. This configuration thus allows us to control sound
radiation in a sensitive way.

%\begin{equation}
%i\hbar\frac{\partial \psi}{\partial t}=
%-\frac{\hbar^2}{2m} \nabla^2 \psi + V\psi + g \vert \psi \vert^2 \psi
%-\mu \psi,
%\label{GP}
%\end{equation}

%\noindent
%where $\psi$ is the macroscopic order parameter of the system, $m$ is the
%atomic mass,  $\mu$ the chemical
%potential and $g$ the strength of the repulsive interaction.
Our analysis is based on numerical simulation of the
Gross-Pitaevskii (GP) equation in the quasi-2D limit
\cite{Parker}.
 Fig.~\ref{trap}(a) shows the density profile of the trapped
condensate in the presence of a vortex which is initially located
near the axis of the trap. The vortex precesses around the axis,
due to the Magnus force: the density gradient due to the trapping
potential causes a buoyancy force which induces tangential motion.
What interests us here is the fact that the acceleration of the
vortex induces emission of sound.
%Since the sound excitations have energy
%of the order of $\mu$, the relative magnitude of $V_0$ and $\mu$
%determines whether the sound energy emitted by the vortex remains
%in the dimple (and is re absorbed by the vortex) or leaves it (and
%is radiated away). The quantity of interest is the vortex energy
%$E_v$, which is obtained by integrating the functional
%$E=-(\hbar^2/(2m)) \vert \nabla \psi \vert^2 + V \psi^2 +
%(g/2)\vert \psi \vert^4$ over a small region around the vortex
%(typically a circle of radius equal to $5\xi$, where $\xi$ is the
%healing length) and subtracting the contribution from the
%background fluid.

\begin{figure}
\centering \includegraphics[height=1.9in,angle=0]{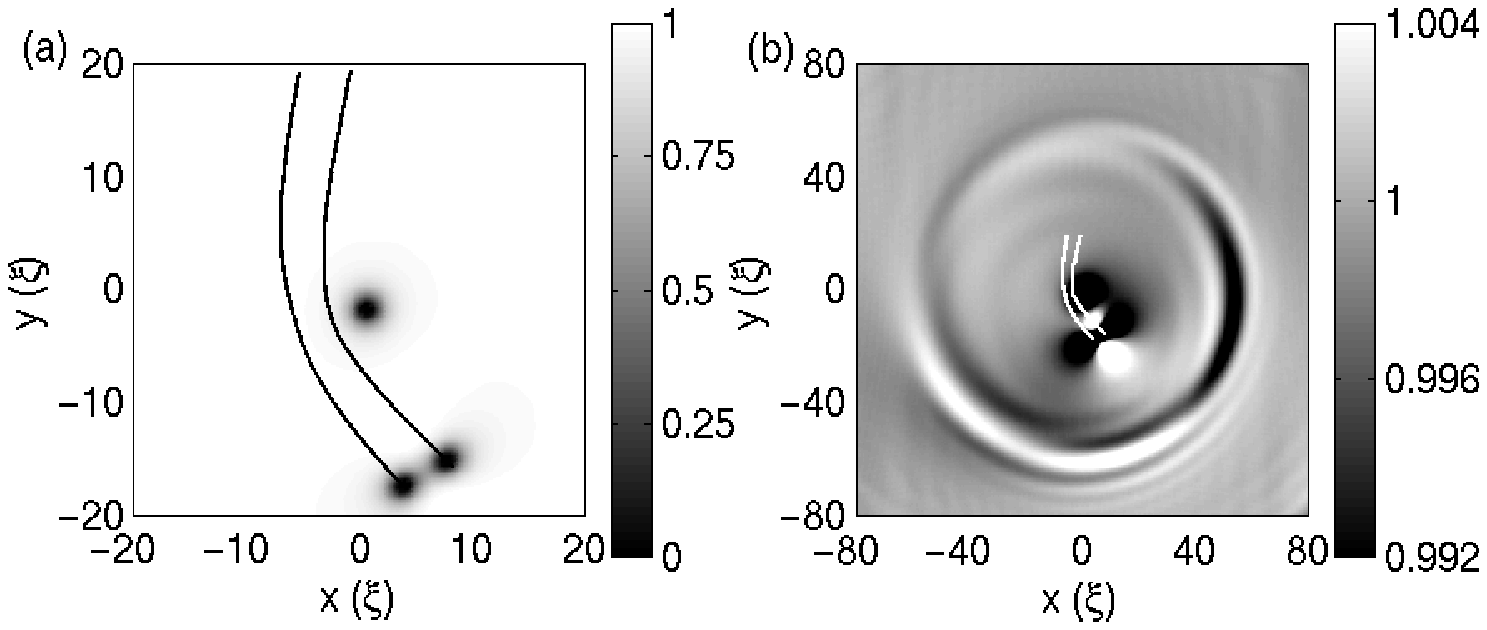}
\centering \includegraphics[height=4.2in,angle=-90]{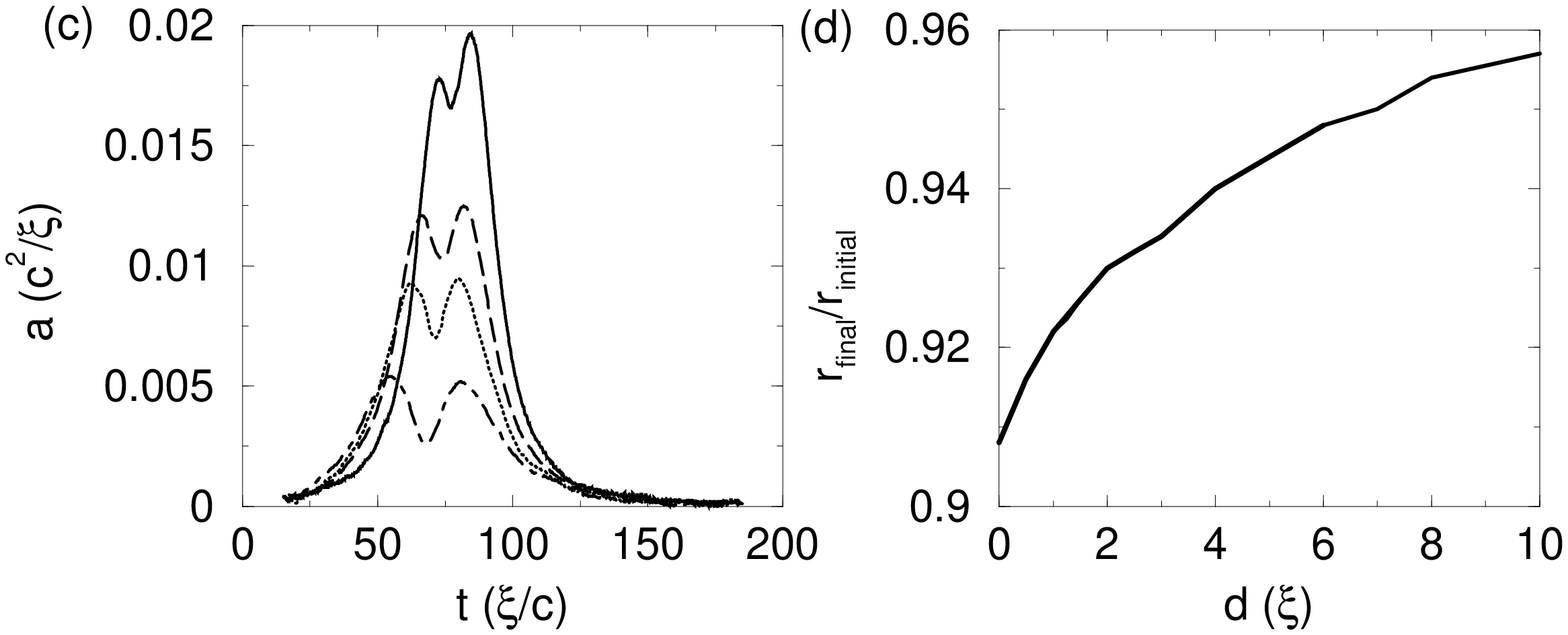}
\caption{Sound burst produced by the close approach of three
vortices. (a) The trajectory of the vortex-antivortex pair (solid
line), initially located at $(0\xi,20\xi)$ and $(5\xi,20\xi)$, is
deflected as it approaches a single vortex located at the origin.
The density plot shows the final density distribution of the
vortices (corresponding to dark spots). (b) The final density
distribution, shown on a different density and length scale, shows
a burst of sound which propagates radially outwards. (c)
Acceleration experienced by the vortex in the pair nearest to the
single vortex, for a vortex pair initially located at $(d,20\xi)$
and $(d+5\xi,20\xi)$, where $d/\xi=0$ (solid line), $1$ (dashed
line), $2$ (dotted line) and $4$ (dot-dashed line). (d) Final
radius of the pair following the interaction (rescaled by the
initial radius of $5\xi$), as a function of distance $d$.
 } \label{flyby}
\end{figure}

\section{RESULTS}

If $V_0\gg \mu$ (deep dimple) the vortex moves around the trap in
a closed orbit, maintaining a mean distance from the axis. The
sound emitted by the accelerating vortex re-interacts with it and
is absorbed. The vortex energy does not decay but executes small
oscillations which result from the beating of the vortex
precession and the modes of the trapped condensate. If $V_0\ll
\mu$ (shallow dimple), the sound radiated by the accelerating
vortex leaves the dimple,  the vortex energy slowly decreases, and
the vortex spirals outwards to lower densities. The sound waves
are emitted in the direction perpendicular to the instantaneous
direction of motion in the form of a dipolar pattern. The
precessional motion of the vortex converts the dipolar emission
into a spiral wave pattern- see Fig.~\ref{trap}(b). For a quasi-2D
BEC, we find\cite{Parker} that the power radiated by the vortex is
$P=\beta m N (a^2/\omega)$, where $a$ is the vortex acceleration,
$\omega$ is the vortex angular velocity (induced by the trap), $N$
is the number of atoms in the BEC, and the dimensionless parameter
$\beta\approx 6.3$ is determined by fitting.  Our result is in
fair agreement with calculations, based on the analogy to $(2+1)D$
electrodynamics\cite{Lundh} (where vortices and phonons map onto
charges and photons, respectively\cite{Arovas}) and classical
acoustics\cite{Vinen}, which yield $\beta=\pi^2/2$, assuming
perfect circular motion for a point vortex in a homogeneous
system.

In Helium II, the relevant parameter is the decay of vortex line
length $L$, typically given by\cite{Vinen} $dL/dt=4\pi
P/[Nm\kappa^2 \ln(b/\xi)]$, where $\kappa=h/m$, and $b$ is the
average intervortex spacing in the tangle. Assuming a
generalisation of our single vortex power radiation formula to a
system of vortices with average separation $L^{-1/2}$, we can cast
our result into Vinen's form\cite{Vinen} $dL/dt=-\alpha
(\kappa/2\pi)L^2$, where $\alpha=\beta/[\pi \ln(\xi L^{1/2})]$. This
generalisation of a numerically-computed expression based on the
GP equation yields the correct power law decay of vortex line
length. Using typical Helium II numbers, we obtain $\alpha \sim
O(0.1)$ for the scenario considered here corresponding to a vortex
decay driven by density inhomogeneity. This is larger than the
value obtained for the `free' decay arising from Kelvin wave
excitations \cite{Leadbeater2}, but still smaller than the
experimentally obtained value for Helium II \cite{Vinen}.  To
establish a link between single vortices driven by trap
inhomogeneity and superfluid turbulence we first consider the
simple example of a vortex-vortex pair. As the vortices corotate
about their central point, they radiate quadrupolar sound waves,
which form a double-armed spiral wave pattern, as shown in figure
Fig.~\ref{pair}(a). Fig.~\ref{pair}(b) shows the acceleration of a
vortex-vortex pair in a homogeneous system and a single vortex in
a harmonic trap. Although the acceleration is similar in both
cases, the sound emission from the pair is less due to the
quadrupolar character of the emission. This suggests that our
value for $\alpha$ is an over-estimate for the case of a vortex
tangle. The acceleration of the vortex-vortex pair tails off as
the vortex cores merge ($r\rightarrow 0$), in contrast to the
analytic prediction of Pismen\cite{Pismen}. Note that the
acceleration of a vortex in a harmonic trap is strongly dependent
on the trap frequency.

As an example of a three-vortex interaction, we consider the close
approach of a vortex-antivortex pair towards a single vortex, as
illustrated in Fig.~\ref{flyby}(a). The single vortex and its
nearest neighbour in the pair have the same polarity.  During this
interaction, the trajectory of the pair becomes deflected (black
lines in Fig.~\ref{flyby}(a)), while the single vortex makes small
deviations about the origin (not shown). The sharp acceleration of
the vortices (indicated in Fig.~\ref{flyby}(c)) induces a sound
burst, which propagates radially outwards. This is visible in the
final density distribution shown in Fig.~\ref{flyby} (b). This
energy loss results in a reduction in the size of the vortex pair,
shown in Fig.\ref{flyby}(d). The magnitude of the loss decreases
as the initial position of the pair is moved further away from the
single vortex, in line with the decreasing acceleration
experienced by the vortices (Fig.\ref{flyby}(c)). Note that the
magnitude of the acceleration induced by the close approach is
similar to the values studied in the dimple trap. Finally, if the
single vortex and its nearest neighbour in the pair have opposing
polarity, the interaction tends to involve a reconnection whereby
these vortices form a pair, and leave behind the other vortex.

In summary, we have shown how sound emission can be controlled and
quantified in experiments on dilute Bose-Einstein condensates.  We
discuss how the information gained can be mapped onto the
understanding of the decay of superfluid turbulence in the limit
of low temperature.  However, further work is needed to exactly
quantify the sound emission for the case of many vortices.
%\begin{figure}
%\centering
%\includegraphics[height=3.0in,angle=-90]{pair2.eps}
%\caption{(a) Acceleration and velocity (inset) of a corotating
%pair as a function of the pair separation $r$, from the time-dependent
%GPE (points), a time-independent numerical approach in the
%rotating frame (solid line), and the analytic prediction
%corresponding to $a=\kappa^2/r^3$ \cite{Pismen}. (b) Acceleration
%and velocity (inset) of a single vortex precessing in a tight
%harmonic trap $\omega=\sqrt{2} \times 10^{-1} (\mu/\hbar)$, from
%the time-dependent GPE. The results are strongly dependent on the
%trap frequency employed.} \label{pair2}
%\end{figure}

\section*{ACKNOWLEDGMENTS}
We acknowledge Mark Leadbeater for data included in
Fig.~\ref{pair} and financial support from EPSRC.

\end{document}